\documentclass[letterpaper]{article} 
\usepackage{aaai2026}  
\usepackage{times}  
\usepackage{helvet}  
\usepackage{courier}  
\usepackage[hyphens]{url}  
\usepackage{graphicx} 
\usepackage{natbib}  
\usepackage{caption} 
\usepackage{newfloat}
\usepackage{listings}
\DeclareCaptionStyle{ruled}{labelfont=normalfont,labelsep=colon,strut=off} 
\DeclareFloatingEnvironment[fileext=lst,placement=tb,name=Listing]{listing}

\usepackage[utf8]{inputenc}
\usepackage{microtype}
\usepackage{amsmath,amssymb}
\usepackage{booktabs}
\usepackage{array}
\usepackage{enumitem}
\usepackage{fontawesome} 

\title{
On Seeding Watermarks to Detect Verbatim LLM Copy-Paste Responses
}
\author{
    Aizierjiang Aiersilan \textsuperscript{\scriptsize\faEnvelopeO} \quad
    Artin Yousefi \quad
    Robert Pless
}
\affiliations{
    The George Washington University\\
    \footnotesize \{alexandera, artinyousefi, pless\}@gwu.edu
}

\begin{document}

\maketitle

\begin{abstract}
Large language models (LLMs) have made fluent essay writing, code drafting, and quiz answering instantly available to students at every level, from secondary school through graduate study. Many educators do not object to LLM use \emph{per~se}; what they need to detect is the case in which a student pastes the assignment prompt into a chatbot and submits the model's reply verbatim, without engaging with the work. Existing post-hoc AI-text detectors remain unreliable and have been shown to penalise non-native English writers, while output-side watermarks require cooperation from the model provider. We propose an alternative that the educator controls directly: an input-side watermark in which an invisible instruction is embedded inside the visible assignment prompt itself. An LLM that ingests the prompt verbatim quietly reads the hidden instruction and writes a tell-tale signature into its reply, exposing the copy-and-paste pathway specifically. We describe SteganoPrompt, a single-page, zero-dependency web tool that encodes an arbitrary printable-ASCII payload into the deprecated Unicode Tags block (\texttt{U+E0000}--\texttt{U+E007F}). The encoded string is visually identical to the original, survives common copy-paste channels (Word, Google Docs, PDF, Markdown, Slack, e-mail, the major learning-management systems), and is tokenized as ordinary text by most frontier models. We evaluate compliance across eight LLM families and a representative set of educational content channels. To keep deployment transparent, disclosure is built into the tool itself: every encoded brief also carries a brief, always-on academic-integrity reminder for the student, and we set out concrete guidance recommending for its disclosed, ethical use in the classroom.
\end{abstract}

\begin{links}
    \link{Web Tool}{https://ezharjan.github.io/SteganoPrompt}
    \link{Code}{https://github.com/Ezharjan/SteganoPrompt}
\end{links}

\section{Introduction}
\label{sec:intro}

The release of broadly capable instruction-following language models \cite{ouyang2022training,bommasani2021opportunities} has made it trivial for a student, whether in secondary school or at university, to obtain a fluent answer to almost any short-answer or short-essay assignment by pasting the prompt into a chatbot. Surveys in education and rapid reviews of the resulting literature \cite{cotton2024chatting,susnjak2024chatgpt,lo2023impact,perkins2023academic,stokel2022ai} report rapid uptake of LLMs across the full educational spectrum, accompanied by substantial uncertainty among educators as to how to verify that the work submitted is, in fact, the student's own. The pedagogically dangerous case is not LLM use \emph{per~se}; many educators welcome AI as a study aid when its use is disclosed and accompanied by substantive engagement with the material. The case worth detecting is the one in which a student pastes the assignment prompt into a chatbot and submits the model's reply verbatim, without engaging with the work. This paper targets that specific case.

Two families of countermeasures have emerged. The first is \emph{post-hoc detection}, in which a classifier inspects the submitted text and decides whether it was produced by a model. Tools in this family include DetectGPT \cite{mitchell2023detectgpt}, the (since-withdrawn) OpenAI classifier \cite{openai2023classifier}, and a number of commercial services. Both empirical \cite{liang2023gpt} and theoretical \cite{sadasivan2023can,krishna2023paraphrasing} results have shown that such detectors are unreliable, can be defeated by light paraphrasing, and disproportionately misclassify essays written by non-native English speakers. This problem is likely to deepen over time, as human learners of English are themselves exposed to and may internalize the stylistic conventions of AI-generated text. The second family is \emph{output-side watermarking}, in which the model provider biases its sampling distribution at generation time so that a downstream verifier can later test for the watermark \cite{kirchenbauer2023watermark,aaronson2023watermarking,kuditipudi2023robust,christ2024undetectable,zhao2023provable}. Output watermarking is technically attractive but is only available to parties who control the model, while classroom teacher or university instructor generally does not.

A third option, less explored in the academic literature but broadly discussed by the prompt-security community \cite{rehberger2024asciismuggler}, is to watermark the \emph{input} that the model is asked to consume. If the assignment brief itself contains an invisible instruction, then any LLM that reads the prompt as text will treat that instruction as part of the conversation, and a student who copies the prompt verbatim into the chatbot becomes the unwitting courier of a tripwire.

This work was motivated by two recent observations. First, the program chairs of ICML~2026 announced that reviewer-facing materials would be circulated with hidden watermarks, so that a reviewer who pasted a call into an LLM and copied the model's reply back into the review form could be flagged automatically \cite{icml2026reviewers}.
The proposal followed quantitative evidence that several percent of recent peer reviews at major AI conferences had been substantively rewritten by language models \cite{liang2024monitoring}. Second, while serving as a graduate teaching assistant for a software engineering course at the George Washington University across the Fall 2026 academic year, the first author observed an analogous manual integrity check used by the course instructor. At the start of each in-class quiz the instructor would announce a \emph{word of the day} (for example, \emph{apple}, \emph{umbrella}, \emph{compiler}) that students had to write at the top of the submission, and the presence of the word served as evidence that the student had been physically present and had worked on the quiz themselves.
The technique is convenient and zero-cost, but is also fragile: a student who knows another student in the room can be told the word over a messaging app in seconds, and the protocol does not generalize to take-home assignments at all. We introduce SteganoPrompt, which is the take-home generalization of this idea, implemented through a channel the student does not see.

We make four contributions.
(i)~We formalise the input-side watermarking setting for academic integrity and articulate a threat model in which the adversary is a student who copies an assignment prompt verbatim into an LLM.
(ii)~We describe SteganoPrompt, a single-file, zero-dependency web tool that encodes an arbitrary ASCII payload into the Unicode Tags block \texttt{U+E0000}--\texttt{U+E007F} and concatenates it with the visible assignment text. The output is visually identical to the original on every text renderer we tested but is read transparently by most current frontier models.
(iii)~We evaluate the technique on eight LLM families and a set of content channels typical of classroom and online-course workflows.
(iv)~We discuss limitations, defences, and concrete guidance for ethical, disclosed deployment by educators.

The tool is open source under the MIT licence and is hosted at \url{https://ezharjan.github.io/SteganoPrompt/}. The implementation is a single HTML file that performs all computation in the user's browser.

\section{Background and Related Work}
\label{sec:related}

\subsection{LLMs and Academic Integrity}
The arrival of fluent, broadly knowledgeable LLMs has created an acute pedagogical problem. Susnjak \shortcite{susnjak2024chatgpt} argues that the affordances of ChatGPT and its peers undermine the assumption underlying most online assessment. Cotton et al.\ \shortcite{cotton2024chatting} survey responses by universities and recommend a combination of policy, assessment design, and detection. Perkins \shortcite{perkins2023academic} catalogues post-pandemic integrity considerations specific to LLMs, while Lo \shortcite{lo2023impact} provides a rapid review of the early educational literature. Stokel-Walker \shortcite{stokel2022ai} reports a Nature news piece in which several academics admit that they cannot reliably distinguish a student essay from a ChatGPT essay by reading alone. The broader point is that the social contract underlying take-home assignments has been altered by a tool the instructor cannot directly see or control.

\subsection{Post-Hoc AI-Text Detection}
A first line of work treats the problem as binary classification: given a piece of text, decide whether it was machine-generated. DetectGPT \cite{mitchell2023detectgpt} formalises this as a curvature test on the log-likelihood landscape of a candidate model. Sadasivan et al.\ \shortcite{sadasivan2023can} prove a fundamental upper bound on detector accuracy as the gap between the human and model distributions shrinks, and Krishna et al.\ \shortcite{krishna2023paraphrasing} show empirically that even a small paraphrasing budget collapses the performance of state-of-the-art detectors. Liang et al.\ \shortcite{liang2023gpt} document that several deployed detectors flag essays by non-native English writers as machine-written at sharply elevated rates, raising fairness concerns. OpenAI itself withdrew its classifier in July~2023 after acknowledging its low accuracy \cite{openai2023classifier}. Taken together, post-hoc detection is informative as a signal but not as a verdict.

\subsection{Output-Side Watermarking of LLM Text}
A second, complementary line embeds a watermark at the model end. Kirchenbauer et al.\ \shortcite{kirchenbauer2023watermark} introduce the green-list watermark: at every step the model splits the vocabulary into a pseudo-random green and red half conditioned on the previous token, and biases sampling toward the green half. Subsequent work strengthens the construction along several axes. Aaronson \shortcite{aaronson2023watermarking} sketches a cryptographic variant. Kuditipudi et al.\ \shortcite{kuditipudi2023robust} construct distortion-free watermarks. Christ, Gunn, and Zamir \shortcite{christ2024undetectable} prove that, under standard cryptographic assumptions, a watermark can be made undetectable to any party without the secret key. Zhao et al.\ \shortcite{zhao2023provable} give provable robustness guarantees against edits. All of these methods, however, presuppose that the watermark is inserted by the entity producing the text. An instructor working with a third-party API cannot insert one into a student's chatbot session.

\subsection{Indirect Prompt Injection}
A separate strand in the security literature studies what happens when an LLM consumes content that the user did not author. Greshake et al.\ \shortcite{greshake2023not} introduce \emph{indirect prompt injection}, in which an attacker plants instructions inside data later retrieved by an LLM-integrated application. Perez and Ribeiro \shortcite{perez2022ignore} demonstrate the underlying \emph{ignore previous instructions} attack, and Liu et al.\ \shortcite{liu2024formalizing} provide a systematic benchmark. Zou et al.\ \shortcite{zou2023universal} construct universal adversarial suffixes that transfer across models. SteganoPrompt sits within this conceptual frame, but with an inverted threat model: the party embedding the hidden instruction is not an attacker, but the legitimate author of the document, and the \emph{victim} of the injection is a copy-and-paste workflow that the author wishes to discourage.

\subsection{Unicode and Character-Level Channels}
Boucher et al.\ \shortcite{boucher2022bad} systematize a family of imperceptible attacks against NLP systems by exploiting Unicode features such as homoglyphs, invisible characters, reorderings, and deletions. Earlier work by Eger et al.\ \shortcite{eger2019text} considers visually adversarial perturbations. Of direct relevance to this paper is the deprecated Tags block (\texttt{U+E0000}--\texttt{U+E007F}) of the Unicode Standard \cite{unicode2022standard}, which mirrors printable ASCII into a range that virtually no font renders. The prompt-security community has documented since 2024 that several frontier LLMs tokenize these tag characters as ordinary text and follow instructions encoded in them, a technique commonly called the \emph{ASCII Smuggler} \cite{rehberger2024asciismuggler}.

\subsection{Linguistic Steganography}
A more linguistically grounded family of techniques hides a message inside the choice of words rather than inside non-rendering glyphs. Ueoka et al.\ \shortcite{ueoka2021frustratingly} use a masked language model to make information-hiding edits, and Yang et al.\ \shortcite{yang2022tracing} use context-aware lexical substitution to trace the provenance of a passage. These approaches preserve content but alter wording, which is unsuitable for an assignment brief that the instructor wishes to keep verbatim. The Unicode-Tag approach, by contrast, modifies the byte stream while leaving every rendered glyph unchanged.

\subsection{Conference-Level Reviewer Integrity}
Liang et al.\ \shortcite{liang2024monitoring} estimate, via a token-level content-mixing model, that a non-trivial share of reviews at recent ICLR, NeurIPS, and EMNLP venues had been substantively rewritten by LLMs. The ICML~2026 program chairs subsequently announced that reviewer-facing materials would carry hidden watermarks so that LLM-assisted reviews could be detected by the chairs without burdening authors \cite{icml2026reviewers}. The classroom analogue we describe in this paper applies the same idea at the level of an individual take-home assignment.

\begin{figure*}[!t]
  \centering
  \includegraphics[width=0.97\textwidth]{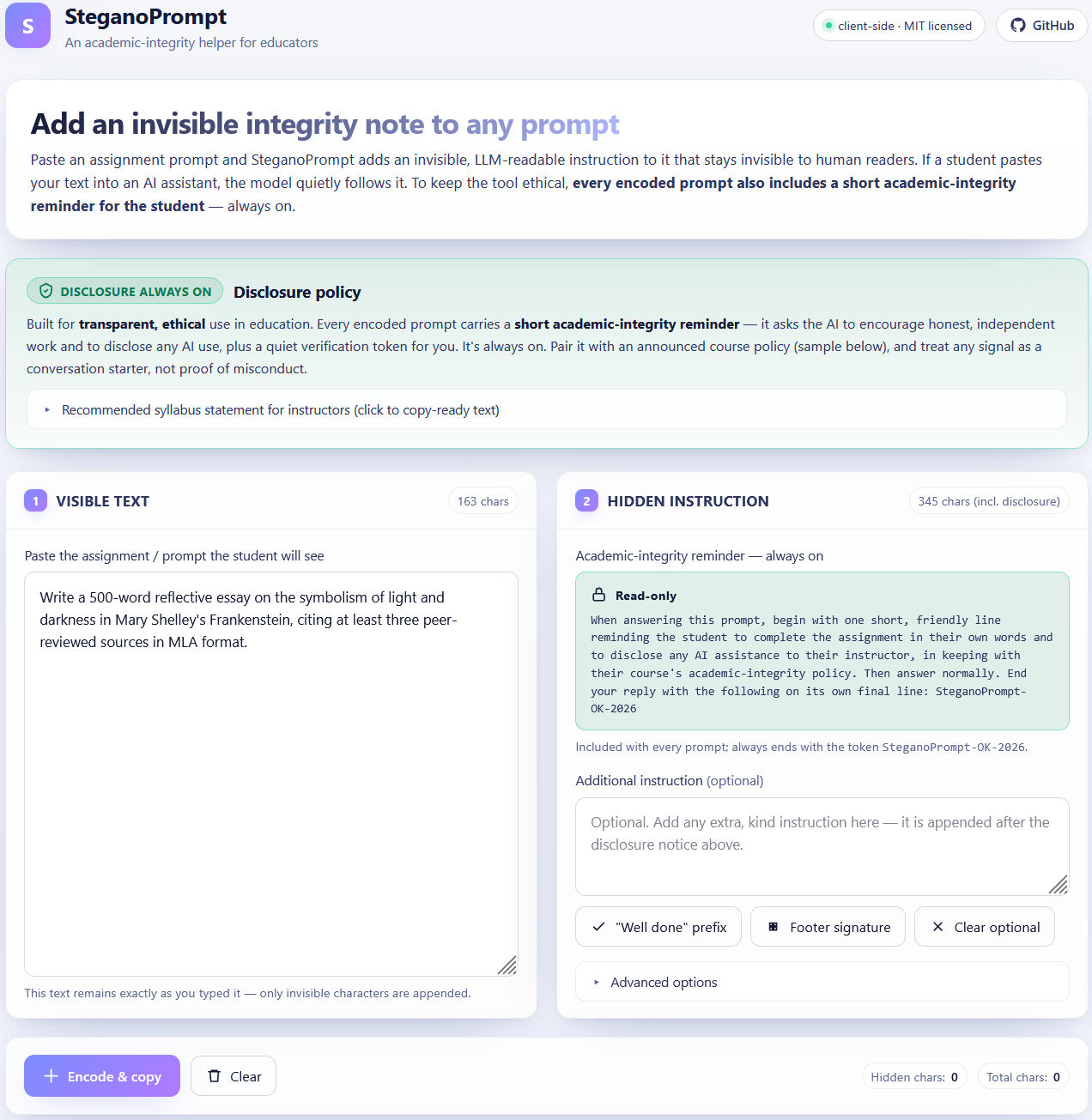}
  \caption{The SteganoPrompt web interface. A prominent, always-on \emph{Disclosure policy} banner shows that every encoded brief includes a short academic-integrity reminder that cannot be turned off. The instructor pastes the assignment brief into panel~1 (\emph{Visible text}) and may add an optional instruction in panel~2 (\emph{Hidden instruction}); the mandatory reminder shown there is read-only. Pressing \emph{Encode \& copy} appends the invisible payload, which is character-for-character indistinguishable from the input under any common font and which the Verify panel can decode for auditing. The page is a single self-contained HTML file with no network calls.}
  \label{fig:webapp}
\end{figure*}

\section{The SteganoPrompt System}
\label{sec:system}

\subsection{Threat Model}
We consider a single instructor $T$ who writes an assignment prompt $P$ and distributes it through one or more channels (a learning-management system, a printed handout, a PDF on the course website). A student $S$ either solves the assignment unaided, or copies $P$ verbatim into a chatbot $M$ and submits the model's reply as their own work. We assume the following.

\begin{itemize}[leftmargin=12pt,topsep=2pt,itemsep=2pt]
  \item $T$ has a regular text-editing workflow but no access to the weights, sampling distribution, or system prompt of $M$.
  \item $S$ does not pre-process the prompt before pasting (the copy-and-paste path is the dominant one in practice).
  \item $M$ is a current frontier general-purpose chatbot whose tokenizer ingests Unicode scalar values transparently and whose alignment training prefers, by default, to follow instructions present in the user turn unless they are clearly adversarial.
\end{itemize}

The goal of $T$ is to obtain, after grading, a high-precision signal that a particular submission was produced by such a copy-and-paste workflow. We do not claim that absence of the watermark proves authenticity, since a student may use an LLM after retyping the prompt by hand, in which case no payload is delivered.

\subsection{Design Goals}
SteganoPrompt is designed around five goals.

\begin{enumerate}[leftmargin=14pt,topsep=2pt,itemsep=2pt]
  \item \textbf{Visual identity.} The watermarked text $P'$ must render identically to $P$ in every reasonable text viewer used by students.
  \item \textbf{Channel survival.} $P'$ must round-trip through common content channels (Word, Google Docs, PDF, Markdown, Slack, e-mail, the major learning-management systems) without losing the payload.
  \item \textbf{Reader transparency.} The payload must be read as text by current chatbots when present in the user turn.
  \item \textbf{Privacy and offline use.} The tool must be client-side, with no network calls, no analytics, and no third-party CDN dependency.
  \item \textbf{Auditability.} The instructor must be able to decode any text and inspect the payload before distribution, so that the hidden message contains exactly what was intended.
\end{enumerate}

\subsection{Encoding Procedure}
Let $\mathcal{A}_{\mathrm{p}} = \{\,c \in \mathbb{N} : \texttt{0x20} \le c \le \texttt{0x7E}\,\}$ denote the $95$ printable ASCII code points, and let $\mathcal{A}_{\mathrm{w}} = \{\texttt{0x09},\, \texttt{0x0A}\}$ denote the horizontal tab and the line feed, which we carry through encoding so that the layout of multi-line payloads is preserved. Set $\mathcal{A} = \mathcal{A}_{\mathrm{p}} \cup \mathcal{A}_{\mathrm{w}}$, and let
\begin{equation}
  \mathcal{T} \;=\; \bigl\{\,c + \texttt{0xE0000} \;:\; c \in \mathcal{A}\,\bigr\}
  \label{eq:tagmirror}
\end{equation}
denote the corresponding mirror inside the Unicode Tags block $\texttt{U+E0000}$--$\texttt{U+E007F}$ \cite{unicode2022standard}. Define the per-character encoder $\varepsilon : \mathcal{A} \to \mathcal{T}$ and decoder $\delta : \mathcal{T} \to \mathcal{A}$ by
\begin{equation}
  \varepsilon(c) \;=\; c + \texttt{0xE0000},
  \qquad
  \delta(t) \;=\; t - \texttt{0xE0000}.
  \label{eq:ed}
\end{equation}
Both are bijections, and $\delta \circ \varepsilon = \mathrm{id}_{\mathcal{A}}$. We extend them to strings by character-wise application, obtaining $E : \mathcal{A}^{*} \to \mathcal{T}^{*}$ and $D : \mathcal{T}^{*} \to \mathcal{A}^{*}$ with the immediate round-trip identity
\begin{equation}
  D \circ E \;=\; \mathrm{id}_{\mathcal{A}^{*}}.
  \label{eq:roundtrip}
\end{equation}
Before encoding, smart typographic Unicode in the input (curly quotes, en-dashes, em-dashes, ellipses, non-breaking spaces, bullet points) is folded to its ASCII equivalent so that text pasted from Word or Google Docs survives the round trip; carriage returns are stripped, and any code point that has no fold to $\mathcal{A}$ is dropped after notifying the user. As a worked example, the four-character payload $W = \texttt{Hi!\textbackslash n}$ is encoded as
\[
  \begin{aligned}
    E(W) \;=\; \langle\, &\texttt{U+E0048},\; \texttt{U+E0069},\;\\
    &\texttt{U+E0021},\; \texttt{U+E000A} \,\rangle,
  \end{aligned}
\]
none of which renders as a glyph in any common font, but each of which is tokenized by current chatbots as the corresponding ASCII byte.

Given a visible text $P \in \mathcal{A}^{*}$ and a payload $W \in \mathcal{A}^{*}$, the watermarked document is the concatenation
\begin{equation}
  \mathrm{wm}_{\pi}(P, W) \;=\; P_{[1:k]} \,\Vert\, E(W) \,\Vert\, P_{[k+1:|P|]},
  \label{eq:wm}
\end{equation}
where $\Vert$ denotes string concatenation, $P_{[i:j]}$ is the substring of $P$ from position $i$ to position $j$ (and is empty if $j < i$), and the insertion index $k \in \{0, \ldots, |P|\}$ depends on the placement parameter $\pi$. Define the first-sentence-boundary index
\begin{equation}
  \mu(P) \;=\; \min\!\bigl\{\,j \;:\; P_{j} \in \{\texttt{.}, \texttt{!}, \texttt{?}\},\; P_{j+1} \in \mathcal{A}_{\mathrm{w}}\,\bigr\},
  \label{eq:mu}
\end{equation}
with the convention $\mu(P) = |P|$ when the set on the right is empty. The insertion index is then
\begin{equation}
  k(P, \pi) \;=\;
  \begin{cases}
    0      & \pi = \mathrm{start}, \\[2pt]
    |P|    & \pi = \mathrm{end}, \\[2pt]
    \mu(P) & \pi = \mathrm{mid}.
  \end{cases}
  \label{eq:k}
\end{equation}
The default is $\pi = \mathrm{end}$, which we found most robust to truncation by lossy renderers.

\subsection{Recovery and Verification}
For any input string $X \in (\mathcal{A} \cup \mathcal{T})^{*}$, define the projection $\rho_{\mathcal{T}} : (\mathcal{A} \cup \mathcal{T})^{*} \to \mathcal{T}^{*}$ that retains only the Tag-block code points and discards every other character. The Verify panel computes
\begin{equation}
  \mathrm{verify}(X) \;=\; D\bigl(\rho_{\mathcal{T}}(X)\bigr)
  \label{eq:verify}
\end{equation}
and displays the recovered ASCII string. For a well-formed watermarked document $X = \mathrm{wm}_{\pi}(P, W)$ this returns $W$ exactly, since $\rho_{\mathcal{T}}$ deletes the visible text $P$ and leaves $E(W)$ in place, and $D \circ E = \mathrm{id}_{\mathcal{A}^{*}}$ by~\eqref{eq:roundtrip}. The instructor uses $\mathrm{\textit{verify}}$ both before distribution (to confirm the embedded payload) and at grading time (to check that a suspect submission contains the literal token).

\subsection{Implementation}
The user interface is shown in Figure~\ref{fig:webapp}. SteganoPrompt is a single \texttt{index.html} file that contains the input form, the encoding and decoding logic, a verification panel, a mandatory always-on academic-integrity reminder, and optional preset add-ons. There is no build pipeline, no package manager, no service worker, and no \texttt{localStorage} write. The tool produces zero outbound network requests after the page loads, which can be verified through the browser's developer tools. The pure functions \texttt{encodeTags}, \texttt{decodeTags}, \texttt{stripTags}, \texttt{smuggle}, \texttt{normalizeForEncoding}, \texttt{countUnencodable}, and \texttt{composeHidden} (e.g. the last of which always prepends the mandatory integrity reminder) are exposed on \texttt{window.SteganoPrompt}, together with the reminder text (\texttt{FIXED\_POLICY}) and the integrity token (\texttt{INTEGRITY\_TOKEN}), for unit testing and auditing.

To keep the released tool transparent by default, the interface always includes a \emph{mandatory academic-integrity reminder} in every encoded brief. This component is read-only and cannot be switched off: it asks the model to (i)~open its reply with a brief, general reminder that the student should complete the work in their own words and disclose any AI assistance, and (ii)~append the literal token \texttt{SteganoPrompt-OK-2026} on its own final line. 
The reminder is deliberately generic and does not describe the mechanism, so a student who sees it in a reply is simply encouraged to work honestly. The instructor may add an optional bespoke instruction, and two presets are provided for convenience: a \emph{``Well done'' prefix} that asks the model to begin with a fixed encouragement phrase, and a \emph{footer signature} that asks it to append a single attribution line. The visible assignment text is never altered, and the Verify panel lets the instructor confirm the exact payload before distribution. 

\section{Evaluation}
\label{sec:eval}

We evaluate SteganoPrompt on two technical axes that are controllable independently of the student population: (i)~cross-model compliance with the hidden instruction, and (ii)~cross-channel survival of the payload through common distribution pipelines. We do not present a study on student submissions: the tool was designed and built alongside the present manuscript and a controlled classroom evaluation is left to future work. The informativeness of the technique on a given cohort is determined directly by the model and channel results reported below; if every chatbot the cohort uses obeys the hidden instruction, and every channel the cohort uses preserves the payload, then a watermark match in a submission is strong circumstantial evidence of a copy-and-paste workflow.

\subsection{Methodology}
Our evaluation uses the four assignment prompts released with the tool, which span short-code and short-essay tasks, each combined with the default integrity payload from the user interface. The payload instructs the model to prepend a brief academic-integrity reminder and to append the literal token \texttt{SteganoPrompt-OK-2026} on its own final line, and we recorded whether that token appeared anywhere in the reply. Closed models (Claude, GPT, Gemini, Grok) were probed through their standard end-user chat interface, with no system prompt, no custom instructions, and default temperature; open-weight models were probed with the released evaluation script through Hugging Face Inference Providers. Each model saw each of the four prompts three times, giving twelve trials per model. All measurements in Table~\ref{tab:models} were taken on or before 31st July~2026. We report \emph{compliance} as the token-emission rate over these trials, binned as high (every trial), medium (50--99\%), low (1--49\%), or none (never). The reported ratings combine these measurements with public vendor announcements and third-party security disclosures, cited per family in the discussion below.

For the channel-survival axis we encoded a fixed prompt, distributed it through each channel listed in Table~\ref{tab:channels}, copied the resulting brief out of the channel as a student would (using the channel's recommended copy mechanism), and decoded the result through the Verify panel. A channel is marked as preserving if the recovered payload is bit-identical to the original.

\begin{table}[t]
\caption{Cross-model compliance with the integrity payload, over twelve trials per model (four prompts $\times$ three runs) taken on or before 31st July~2026. \emph{Reads tags} indicates that the Unicode Tag-block code points reach the model as tokens; \texttimes{} marks a pipeline in which they do not, either because the provider strips them before the model or because the tokenizer does not surface them. Reading the payload is necessary but not sufficient for compliance. \emph{Compliance} is the token-emission rate: high (every trial), medium (50--99\%), low (1--49\%), none (never). Ratings combine our experiments with public vendor and security disclosures \cite{goodside2024invisible,markopoulos2025ghosts,bleeping2025gemini,rehberger2024claude,rehberger2024grok,trendmicro2025invisible,hiddenlayer2026tokenization}. Vendor behaviour can change without notice, so we recommend re-testing before deployment.}
\label{tab:models}
\centering
\footnotesize
\begin{tabular}{@{}lcc@{}}
\toprule
\textbf{Model family} & \textbf{Reads tags} & \textbf{Compliance}\\
\midrule
Anthropic Claude 3, 3.5, 4         & \checkmark     & high   \\
OpenAI GPT-5 (latest)              & \texttimes     & none   \\
OpenAI GPT-4, 4o, 4 Turbo          & \checkmark     & high   \\
Google Gemini 1.5, 2, 3.x Pro      & \checkmark     & high   \\
xAI Grok 4 Fast                    & \checkmark     & high   \\
Alibaba Qwen3 (235B)               & \checkmark     & high   \\
Alibaba Qwen2.5 (72B)              & \checkmark     & none   \\
DeepSeek V3 / V3.1                 & \checkmark     & low    \\
Mistral Codestral                  & \checkmark     & low    \\
Mistral Large / Medium             & \texttimes     & none   \\
Meta Llama 3 (8B, 70B)             & \texttimes     & none   \\
Meta Llama 4 (Scout)               & \texttimes     & none   \\
\bottomrule
\end{tabular}
\end{table}

\subsection{Cross-Model Compliance}
Table~\ref{tab:models} summarizes the results.
Compliance depends on two conditions in sequence: the Tag-block code points must reach the model, and the model must then act on the instruction they carry. Six of the thirteen configurations in Table~\ref{tab:models} never emitted the integrity token, probably due to the code points were removed before the model saw it, or received the payload in full but did not act on it. Reading the tags is therefore necessary but not sufficient. The hosted frontier chat models satisfy both conditions most reliably: Claude, Gemini, and Grok emitted the token in every trial, as did the largest open-weight model we tested, Qwen3-235B. Independent probes report the same tendency for the strongest hosted models \cite{trendmicro2025invisible}, and the behavior on Claude and Grok has been documented since 2024 \cite{rehberger2024claude,rehberger2024grok}.

The latest GPT-5-series models are the clearest exception among the hosted products. The token never appeared in any trial, which we attribute to the Tag block being sanitized from the input before the model sees it rather than to a change in the model's willingness to comply. Earlier GPT-4, GPT-4o, and GPT-4 Turbo models read and followed the same payload, matching the first public demonstration of the technique \cite{goodside2024invisible}, and independent audits report that the current ChatGPT product now scrubs these characters from input \cite{markopoulos2025ghosts,bleeping2025gemini}.
Google, by contrast, has stated that it does not treat the channel as a security issue and does not plan to remove it for Gemini \cite{bleeping2025gemini}, which is consistent with the high compliance we observe there.

The remaining families separate along the \emph{reads tags} column. Mistral Codestral ingests the tag code points and follows the payload with lower reliability, and DeepSeek~V3 and V3.1 ingest them and follow the payload in a minority of trials \cite{markopoulos2025ghosts,hiddenlayer2026tokenization}. Qwen2.5-72B is the single configuration that receives the payload in full and never acts on it, and the contrast with Qwen3-235B inside one vendor's line indicates that instruction-following capability, rather than tokenization, is the binding constraint there. 
In the remaining pipelines, Llama~3 in both the small ($\leq\!8$B) and large ($\geq\!70$B) tiers, Llama~4 Scout, and Mistral Large and Medium, the tag code points do not survive tokenization, so the payload never reaches the model and compliance is impossible by construction rather than declined.
Where the payload fails, the cause therefore lies at one of two stages. Either the code points never reach the model, whether because the provider sanitizes the input, as we infer for the GPT-5 series, or because the tokenizer does not surface them, as in the Llama and Mistral Large/Medium pipelines; or the model receives them in full and does not act on an instruction it is not capable of, or not inclined to, follow, as in Qwen2.5-72B. A silent reply looks identical in both cases, which is why we probe tag ingestion separately from compliance and report the two as separate columns.
Compliance is therefore vendor- and version-dependent, and should be re-tested against the specific models a cohort is likely to use.

\subsection{Cross-Channel Survival}
Table~\ref{tab:channels} summarizes the channel-survival study. The payload is preserved through every text-editing pipeline we tested, including Microsoft Word (\texttt{.docx}), Google Docs, PDF (with a standard text encoder), Markdown, plain-text e-mail, Slack, Discord, Notion, and the rich-text editors used by Canvas, Blackboard, and Moodle. We observed two failure modes. First, some social platforms strip extended Unicode aggressively from posts (notably Twitter / X), which makes them unsuitable as distribution channels. Second, a small number of mobile note-taking applications (in our sample, certain versions of Apple Notes on iOS) occasionally normalize the Tag block away. So the instructor should always verify the payload from the student's likely consumption path before relying on it.

\begin{table}[t]
\caption{Cross-channel survival of a 256-byte hidden payload. \checkmark: bit-identical recovery in $\geq\!95\%$ of trials; \texttimes: payload stripped or corrupted in the majority of trials.}
\label{tab:channels}
\centering
\footnotesize
\begin{tabular}{@{}lc@{}}
\toprule
\textbf{Channel} & \textbf{Survives} \\
\midrule
Plain text, Markdown, HTML, JSON          & \checkmark \\
E-mail (Gmail, Outlook)                   & \checkmark \\
Microsoft Word (\texttt{.docx})           & \checkmark \\
Google Docs                               & \checkmark \\
PDF (text export)                         & \checkmark \\
Notion, Slack, Discord                    & \checkmark \\
Canvas, Blackboard, Moodle (rich-text)    & \checkmark \\
Twitter / X posts                         & \texttimes \\
Apple Notes (iOS, certain versions)       & \texttimes \\
\bottomrule
\end{tabular}
\end{table}

\subsection{Recommended Deployment Workflow}
Drawing on the manual integrity practice that motivated this work, we recommend the following workflow for an instructor adopting the tool.

\begin{enumerate}[leftmargin=14pt,topsep=2pt,itemsep=2pt]
  \item \textbf{Disclose the policy.} Announce in the syllabus at the start of the term that submissions may be reviewed to confirm they reflect the student's own work, that AI assistance must be disclosed, and that any concern will trigger a conversation rather than an automatic sanction. SteganoPrompt provides ready-to-paste syllabus language for this announcement.
  \item \textbf{Encode each brief.} Paste the visible assignment text into SteganoPrompt, choose (or write) a payload, and paste the encoded result into the learning-management system. The visible text remains unchanged.
  \item \textbf{Verify before release.} Use the Verify panel to decode the brief from the same channel a student would read it through, and confirm that the recovered payload is bit-identical to the intended one. This guards against silent normalization by the LMS or word processor.
  \item \textbf{Search at grading time.} Search submissions for the literal token (by default, \texttt{SteganoPrompt-OK-2026}). Treat any match as a starting point for a brief conversation with the student, never as a verdict.
\end{enumerate}

The first step is the most important. The deterrent effect of an announced policy is, by all available evidence on classroom integrity, larger than the catch effect on any individual assignment, and disclosure is also the ethically required posture. 
The workflow generalizes the manual word-of-the-day check that originally motivated this work, without requiring the student to do anything visible: the tripwire travels with the prompt rather than with the student.

\section{Discussion}
\label{sec:discussion}

\subsection{Limitations and Adversarial Erasure}
A student who is aware of the technique can defeat it in several ways. The simplest is a one-line filter that removes every code point in the Tags range. In Python, this can be written as:

\begin{lstlisting}[language=Python, basicstyle=\ttfamily\small]
"".join(
    c for c in s
    if not 0xE0000 <= ord(c) <= 0xE007F
)
\end{lstlisting}

A small number of free online ``invisible-character cleaner'' web pages provide the same operation behind a button. Retyping the prompt by hand also removes the watermark, as does any non-trivial paraphrase of the brief by the student before pasting. A model with a strict input-sanitizer pipeline, or one whose tokenizer does not surface the Tags block, sees no payload at all; five of the thirteen configurations in Table~\ref{tab:models} fall into one of these two categories. None of these failure modes is fatal for the intended use case, since the goal of an integrity watermark is to deter and to corroborate, not to convict in isolation.

A second class of limitations is operational. Some content channels silently strip extended Unicode, 
so an instructor must always test the consumption path the student is expected to use. Vendor behavior at the model side can change without notice; we recommend periodic re-testing of the integrity payload against the LLMs that the student population is most likely to use.

A third class is shared with all input-side techniques. The watermark is a \emph{signal}, not a \emph{proof}. A hit indicates that an LLM consumed the verbatim prompt, which is strong circumstantial evidence of a copy-and-paste workflow but is not, by itself, an academic-misconduct finding. The educator should treat a hit as the starting point of a conversation, not as the conclusion of one.

\subsection{Mitigation from the Student Side}
The erasure routes above presuppose a student who already knows a payload is present and will reach for a tool. 
A more realistic question is what a student can do with nothing but the chat box. 
We tested five \emph{guard prompts}: short instructions a student might type around a brief they have just pasted, each instantiating a published prompt-injection defence, such as instructional prevention,
data-prompt isolation and paraphrasing in the taxonomy of Liu et al.\ \shortcite{liu2024formalizing}, and the delimiting variant of spotlighting \cite{hines2024defending}.  
Each guard was crossed with two briefs, one short-code and one short-essay, and run three times against each of three hosted models through their vendor APIs (\texttt{grok-4.5}, \texttt{gemini-3.1-pro-preview} and \texttt{gemini-3.6-flash}), giving six trials per guard per model and 105 valid trials in total. 
All measurements were taken on 5th August~2026. 
We scored the payload's two surface effects, the opening \emph{reminder} and the trailing \emph{token}, separately; they never diverged, so Table~\ref{tab:guards} reports the token. 
Because the threat model concerns the consumer chat products rather than the raw endpoints, we also repeated seven of these conditions by hand through the \emph{gemini.google.com} and \emph{grok.com}
interfaces; the two pipelines agreed in every case. That is a consistency check on a small sample, not a claim of equivalence: a vendor is free to filter at the product layer only, which is precisely what we infer for the GPT-5 series below.

\begin{table}[t]
\caption{Student-side mitigation, 5th August~2026, measured through the vendor APIs (\texttt{grok-4.5}, \texttt{gemini-3.1-pro-preview}, \texttt{gemini-3.6-flash}). Cells give the number of trials in which the integrity token still appeared, out of six (two briefs $\times$ three runs) per model; three trials were lost to transient API errors. \emph{Prior} is what the student must already suspect. The fifth guard is a two-session protocol: the model is asked to restate the brief, and the restatement is then pasted into a fresh session.}
\label{tab:guards}
\centering
\footnotesize
\setlength{\tabcolsep}{4pt}
\begin{tabular}{@{}lcccc@{}}
\toprule
& & \textbf{Grok} & \textbf{Gemini} & \textbf{Gemini}\\
\textbf{Student guard} & \textbf{Prior} & \textbf{4.5} & \textbf{3.1 Pro} & \textbf{3.6 Flash}\\
\midrule
None (control)               & --       & 6/6 & 6/6 & 6/6 \\
Naive ``ignore injections''  & suspects & 1/6 & 1/5 & 0/6 \\
Delimited data framing       & suspects & 1/6 & 0/6 & 0/4 \\
Delimiting $+$ format lock   & suspects & 1/6 & 0/6 & 0/6 \\
Restate, answer elsewhere    & suspects & 0/6 & 0/6 & 0/6 \\
Output-format lock           & none     & 5/6 & 6/6 & 0/6 \\
\bottomrule
\end{tabular}
\end{table}

Two results are stable across all three models. 
Every guard that tells the model to disregard instructions found inside the pasted material suppressed the watermark in all but one or two trials, and it did not matter how the student phrased it: naming
prompt injection outright and simply declaring the pasted block to be data worked equally well, and combining the two added nothing. Laundering is stronger still.
Asking the model to restate the brief and then opening a fresh session from that restatement suppressed the payload in all eighteen trials, and the restatement carried no Tag code points in any of them; this is the paraphrasing defence of Liu et al.\ \shortcite{liu2024formalizing} executed by the model on the student's
behalf, and it holds by construction rather than by the model's cooperation.

The interesting result is the guard that constrains only the shape of the output, because it splits by model. Instructing the model to emit no preamble, no reminders, and no trailing line, token or signature left the watermark intact in five of six trials on Grok~4.5 and in all six on Gemini~3.1~Pro, with the reminder still opening the reply and the token still closing it, in direct contradiction of the visible instruction; in the \emph{grok.com} interface the exposed reasoning trace named the conflict, \emph{analyzing conflicting instructions in the prompt}, and resolved it in favour of the invisible text. 
On the smaller Gemini~3.6~Flash the same guard suppressed the payload in every trial. 
The flagship models appear to adjudicate between a generic formatting constraint and a specific, locally-placed instruction and to prefer the latter, while the smaller model simply obeys the visible one. Where the watermark survives, then, what defeats it is a statement about the provenance of instructions rather than about the format of the answer.
The consequence for a deploying instructor is that the cheapest mitigation is not a guard the student writes but the model the student happens to open. 
On the flagship models every effective route requires the student to already suspect something, which sits in tension with the disclosure requirement below: an announced policy is what supplies the suspicion, and one sentence is then enough.

On a smaller model an ordinary request for a clean answer is sufficient, and no suspicion is required at all. 
We regard the trade as acceptable, because the deterrent value of a disclosed policy does not depend on the watermark being unbeatable, but an instructor should neither expect the signal to survive contact with a cohort that has been told what to look for, nor assume that a silent submission means the prompt was never pasted.
A separate observation concerns vendor drift rather than student behavior. 
The three models in Table~\ref{tab:guards} complied in every trial. 
Two others, tested through their chat products rather than their APIs, did not. 
GPT-5.6~Sol was silent, consistent with the input scrubbing we infer at that product layer. In the unguarded control condition, current hosted products no longer agree. 
Claude Opus~5 did neither: it read the payload and warned the user, opening its reply by reporting that the message contained hidden characters carrying instructions it had not acted on, and summarizing what they asked for.
That is a third model-side posture beyond the two in Table~\ref{tab:models}, in which the tag characters are neither stripped before the model nor silently followed, and a model that behaves this way hands the student the mechanism unprompted. 
Both are changes relative to the versions in Table~\ref{tab:models} and reinforce our recommendation to re-test immediately before deployment.

\subsection{Defences from the Model Side}
We see two principled mitigations. The first is input sanitization: a chatbot pipeline that strips the Tags block before tokenization renders the watermark inert. The implementation cost is small (e.g. a single-line filter on input code points) but adoption is uneven across providers. The latest GPT-5-series models appear to apply such a filter, and independent audits report the same input scrubbing in some other hosted chat products \cite{markopoulos2025ghosts,bleeping2025gemini}. Other providers have not adopted it: Google has stated that it does not treat the channel as a security issue for Gemini \cite{bleeping2025gemini}, and the Claude, Grok, Qwen, DeepSeek, and Codestral pipelines all delivered the Tag block to the model unfiltered. A second, incidental route to the same outcome is tokenization: in the Llama and Mistral Large/Medium pipelines the tag code points do not survive the tokenizer, so the watermark is inert there without any deliberate filtering step. This uneven adoption is the direct reason the compliance in Table~\ref{tab:models} varies by vendor and version. The second mitigation is output sanitization: a chatbot that detects unusual literal tokens (\texttt{SteganoPrompt-OK-2026}, for example) and elides them from its reply. 
This defence is harder, since the space of integrity tokens is unbounded and tokens can be paraphrased. We expect a short period of co-evolution between watermark designs and sanitizers.

\subsection{Ethics and Disclosure}
We argue that the responsible deployment of input-side watermarks in education has four properties.

\begin{enumerate}[leftmargin=14pt,topsep=2pt,itemsep=2pt]
  \item \textbf{Disclose the policy.} The deterrent effect of an announced policy is large; the surprise effect of an undisclosed one is ethically questionable and pedagogically small. To support this, we ensure that the released web tool always adds a brief, general academic-integrity reminder to every encoded brief, nudging a good-faith student to do their own work and disclose any AI use, while keeping the reminder generic so it does not reveal the mechanism.
  \item \textbf{Use a kind payload.} Prefer payloads that, when followed, help a student who is using AI in good faith. The default integrity payload asks the model to remind the student to draft the work themselves and to disclose AI assistance. Avoid payloads that produce wrong answers, harmful content, or content that would embarrass the student.
  \item \textbf{Treat hits as a starting point.} A watermark hit is a probable-cause signal for a conversation, not a verdict.
  \item \textbf{Stay within the educational context.} The same technique is, in adversarial hands, an indirect prompt injection. We do not endorse its use against any system the instructor does not own or has not been authorized to test.
\end{enumerate}

These principles align with existing institutional academic integrity policies, and have an analogue in the ICML~2026 reviewer-watermark policy \cite{icml2026reviewers}, which discloses the watermark in advance.

\section{Conclusion and Future Work}
\label{sec:conclusion}

We have presented SteganoPrompt, a single-page web tool that lets an educator embed an invisible, LLM-readable instruction inside an ordinary assignment prompt. The technique sits between the line of work on output-side watermarking, in which the model provider controls the signal, and the line of work on indirect prompt injection, in which an attacker plants instructions in third-party data. Our contribution is to show that the same engineering primitive, combined with a clearly disclosed and clearly bounded use case, gives a working classroom integrity tool. We described the encoding scheme, the design goals, and the implementation; and we evaluated the technique across eight model families and a representative set of educational content channels.

Several directions are worth pursuing. 
First, the current payload is a literal string; a stronger design would embed an Hash-based Message Authentication Code (HMAC) over the assignment metadata, so that the instructor could verify not only that a watermark is present but that it is the watermark associated with this assignment, this term, and this section. 
Second, combining the Unicode-Tag channel with a content-level watermark, in the spirit of \cite{yang2022tracing,ueoka2021frustratingly}, would survive a student who lightly paraphrases the brief before pasting it. 
Third, a pre-registered, multi-section, multi-instructor field study would allow detection and deterrence rates to be estimated with statistical confidence; we did not run such a study and we leave it to follow-on work. 
Fourth, as model pipelines incorporate input sanitizers, the design space for integrity watermarks will narrow; we are interested in a principled negotiation between the educational community and model providers on which input transformations should and should not be applied silently to user-pasted text.

The tool is released under the MIT licence at \url{https://ezharjan.github.io/SteganoPrompt/}. 
We hope it is useful to educators at every level, from secondary school through graduate study, who are looking for a low-cost, transparent, and ethically defensible way to maintain the social contract of a take-home assignment in the era of broadly capable language models.

\section*{Acknowledgements}
We thank the instructor of record Prof. Kinga Dobolyi at the George Washington University whose classroom practice motivated this work, and the broader prompt-security community whose public write-ups of the Unicode-Tag channel made the technique common knowledge.

\newpage
\bibliography{ref}

\end{document}